\documentclass[fleqn,10pt]{wlscirep}
\usepackage[utf8]{inputenc}
\usepackage[T1]{fontenc}
\title{Leadership emergence in walking groups}

\author[1,2]{Maria Lombardi}
\author[3,*]{William H. Warren}
\author[1,2,*]{Mario di Bernardo}
\affil[1]{Department of Engineering Mathematics, University of Bristol, Bristol, UK}
\affil[2]{Department of Electrical Engineering and Information Technology, University of Naples Federico II, Naples, Italy}
\affil[3]{Department of Cognitive, Linguistic and Psychological Sciences, Brown University, Providence, RI 02912, USA}

\affil[*]{Corresponding authors: bill\_warren@brown.edu, m.dibernardo@bristol.ac.uk}

\begin{abstract}
Understanding the mechanisms underlying the emergence of leadership in multi-agent systems is still under investigation in many areas of research where group coordination is involved. While leadership has been mostly investigated in the case of animal groups, only a few works address the problem of leadership emergence in human ensembles, e.g. pedestrian walking, group dance.
In this paper we study the emergence of leadership in the specific scenario of a small walking group. Our aim is to unveil the main mechanisms emerging in a human group when leader or follower roles are not designated a priori. Two groups of participants were asked to walk together and turn or change speed at self-selected times. Data were analysed using time-dependent cross correlation to infer leader-follower interactions between each pair of group members. The results indicate that leadership emergence is due both to contextual factors, such as an individual's position in the group, and to personal factors, such as an individual's characteristic locomotor behaviour. Our approach can easily be extended to larger groups and other scenarios such as team sports and emergency evacuations.
\end{abstract}

\begin{document}

\flushbottom
\maketitle
\thispagestyle{empty}

\section*{Introduction}
\label{sec:introduction}
A number of interesting phenomena can emerge from the collective behaviour of a group of agents interacting with each other that cannot be explained in terms of the dynamics of an individual agent \cite{Abdallah2007,Bullo2019}. One phenomenon of extreme importance is the emergence of leadership. Leadership plays a crucial role in determining the success or the failure of a variety of activities both in animals (e.g. protecting the animal group against predatory attacks, locating food resources) and human groups (e.g. steering opinion dynamics, playing music or sport).
Many studies exist in the literature addressing coordination and leadership emergence in the case of groups of animals \cite{Reebs2000,Couzin2005,Zienkiewicz2015,Nagy2010,Orange2015}, where the problem is tackled by studying how agents informed about a migration route or food position influence collective behaviour. In this context, results shows that, as group size becomes larger, only a small proportion of informed individuals is needed to guide the group behaviour.

Limited results are available on leadership emergence in human ensembles \cite{DAusilio2012,Rio2014b,Dyer2009}. Examples include consensus decision making \cite{Boos2014,Dyer2008,Dyer2009}, the role of a conductor in an orchestra \cite{DAusilio2012,Glowinski2010}, group behaviour in dance \cite{Ozcimder2016,Leonard2014}, among many others.

Here, we study the emergence of leadership in a group of people walking together. The goal is to assess whether leadership emerges and whether it facilitates group coordination and cohesiveness. Understanding leadership emergence in groups of humans walking together can be relevant in many applications. For example, during evacuation from a building, the presence of a leader with global knowledge of the structure can safely guide people towards the exit avoiding fatal consequences \cite{Pelechano2006}. Past studies have observed that, in these emergency situations, some people have a higher probability of becoming leaders according to their attitude or qualities, e.g. ability to handle emergence situations, charisma, confidence, friendliness, effective speaking \cite{Stogdill1948}. Also, it has been observed that such a leader is characterised by specific behavioural or physiological traits such as a common tendency to act first \cite{King2009}. On the other hand, the spontaneous emergence of leadership may depend on contextual factors, such as an individual's position in the group \cite{Nagy2010}.

To investigate leadership emergence in human groups, we consider the simplest scenario in which four participants are asked to walk together across the room. Our aim is to unveil whether, when no designated roles are assigned, a leader emerges in the group. Specifically, we adopt a time-dependent cross correlation analysis to unfold the leader-follower relationships established within the group by instructing all participants to change their heading direction or speed twice at their will during every trial. We then analyse each trial to identify the individual whose change of locomotion (direction or speed) is readily followed by other group members. That individual is then labelled as the leader on that trial. We further analyse how relative positions in the group, or a participant's characteristic locomotion behaviour, influence leadership emergence.

Our findings confirm the observations made in \cite{Rio2014b,Dachner2014,Rio2014a} that leadership depends on local visual couplings between group members that try to match their speed and heading direction accordingly. Also, our data analysis shows that in some situations a participant's characteristic locomotion behaviour can be so influential as to dominate the group independent of his/her position.

\section*{Results}
\label{sec:results}
Two groups of four participants were tested during the experiment, designated Group $1$ and Group $2$. Before each trial, group members were positioned on the vertices of a square configuration, labelled as FL (front left), FR (front right), BL (back left) and BR (back right) (refer to \emph{Methods}, Figure \ref{fig:group_config}). To identify an individual participant, we use the notation ``P$x.y$" where $x$ is the group and $y$ is a sequential number identifying a specific participant, for example P$1.3$ indicates participant $3$ in group $1$. During each trial, group members were instructed to change direction (turn left/turn right) or speed (speed up/slow down) twice while staying together as they walked across the room. Three different initial inter-personal distances (IPD) between participants were tested: $1$, $2$ and $4$ metres. Participants' positions were shuffled between each trial.

The data were analysed to assess the effect on leadership emergence of (i) spatial position in the group and (ii) individual locomotor behaviour. We compared the time series of heading (or speed) for each participant by adapting the time-dependent cross correlation metric proposed in \cite{Giuggioli2015} to characterise leadership emergence in groups of trawling bats. Briefly, the method quantifies the temporal dependence of turns (or speed changes) between group members, allowing us to compute an \textit{individual leadership index} for each participant on each trial (for further details and formal definition of the index see \emph{Methods}).

\vspace{0.5cm}

\begin{figure}[ht]
	\centering
	\includegraphics[width=\linewidth]{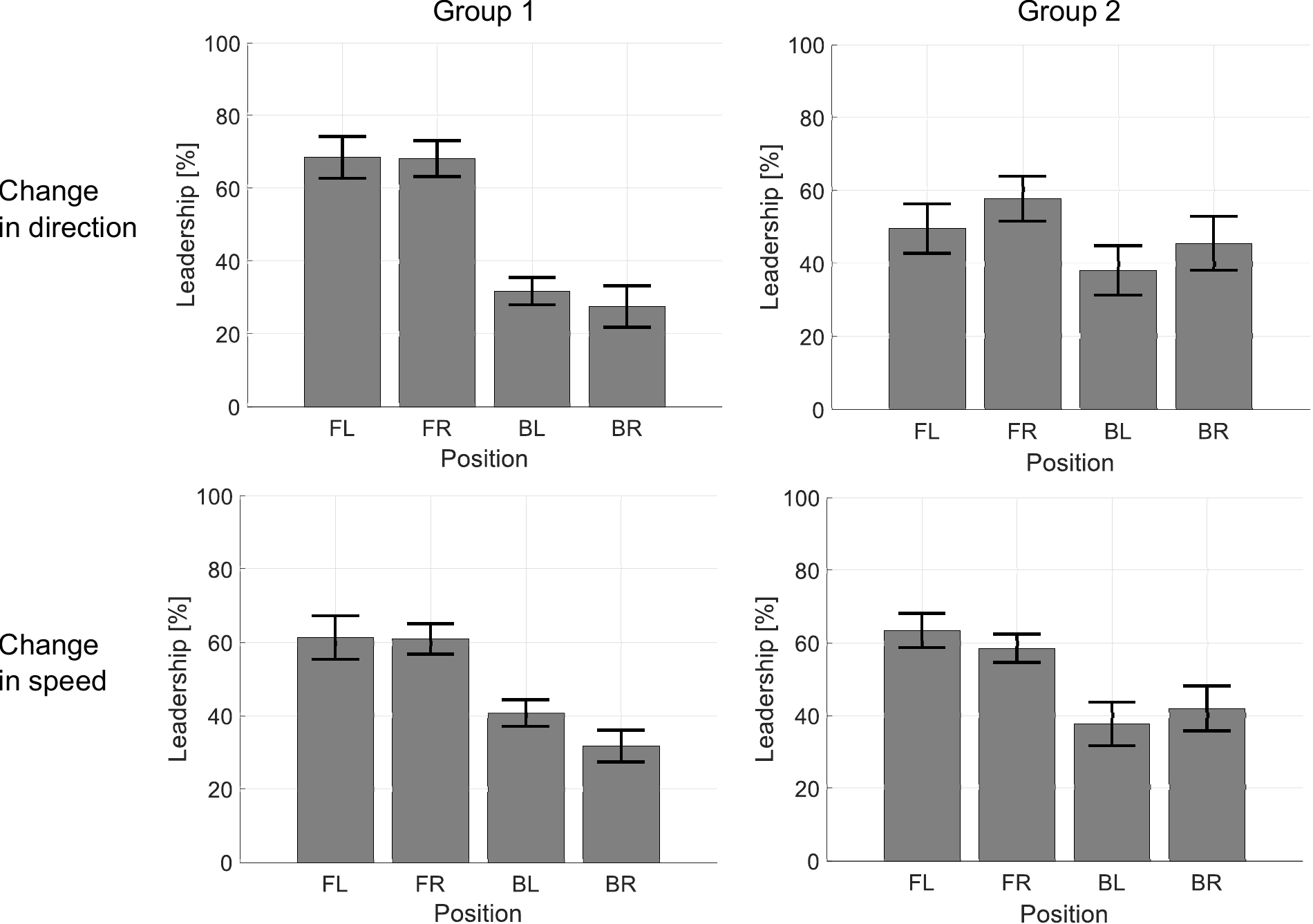}
	\caption{\textbf{Averaged percent leadership evaluated grouping the trials by position}. Four bar charts are reported, $2$ groups $\times$ $2$ condition (speed and heading direction). Each bar represents the mean value of the leadership grouped by position over the number of trials, whereas grey error bars are the standard error of the mean. FL = front left, FR = front right, BL = back left, BR = back right.}
	\label{fig:positions_comparison}
\end{figure}

\subsection*{Effect of pedestrian position in the group} Due to the square configuration used during the experiments, we expected one of the pedestrians in the front row to act as the leader because they can influence participants in the back row (i.e. the back row is visually coupled to the front row), but not vice versa.

Figure \ref{fig:positions_comparison} shows the influence of participants' position on the \textit{individual leadership index}, averaged over all trials, in the two conditions (heading direction and speed). Results indicate that participants in the front left and front right positions have the highest value of the leadership index.

This is confirmed by a formal statistical analysis. A three-way Mixed ANOVA (2 Conditions $\times$ 4 Positions $\times$ 3 IPD) was performed on percent leadership, in which Condition and IPD were treated as ``between-subjec'' factors, Position was treated as a ``within-subject'' factor, and Group was considered a random effect. The analysis revealed a significant main effect of Position on the measured leadership ($F(3,117)=19.865, p < 0.05, \eta^2 = 0.337$).

The Condition main effect and the Condition $\times$ Position interaction were not in themselves significant, having respectively $F(1,39)=1.334, p = 0.255, \eta^2 = 0.033$ and $F(3,117)=0.196, p = 0.899, \eta^2 = 0.005$, indicating a similar influence of position on leadership for heading and speed. Furthermore, neither the IPD main effect ($F(2,39)=0.133, p = 0.876, \eta^2 = 0.007$) nor the IPD $\times$ Position interaction ($F(6,117)=0.906, p = 0.493, \eta^2 = 0.044$) were significant, indicating a similar influence at all distances.. Finally, the three-way Condition $\times$ IPD $\times$ Position interaction was not significant, $F(6,117)=1.301, p = 0.262, \eta^2 = 0.063$. Thus, the factor Position alone dominates percent leadership.

Since the factor Position was statistically significant, we performed Bonferroni post hoc test for a pairwise comparison in order to understand in detail what positions influence most leadership emergence. We found that the differences between the front positions and the back positions are statistically significant ($p < 0.05$) while the differences between the two positions in the front and between the two in the back have a $p=1$ and so not statistically significant. These results demonstrate a clear contextual effect of position on emergent leadership.

\vspace{0.5cm}

\begin{figure}[ht]
	\centering
	\includegraphics[width=\linewidth]{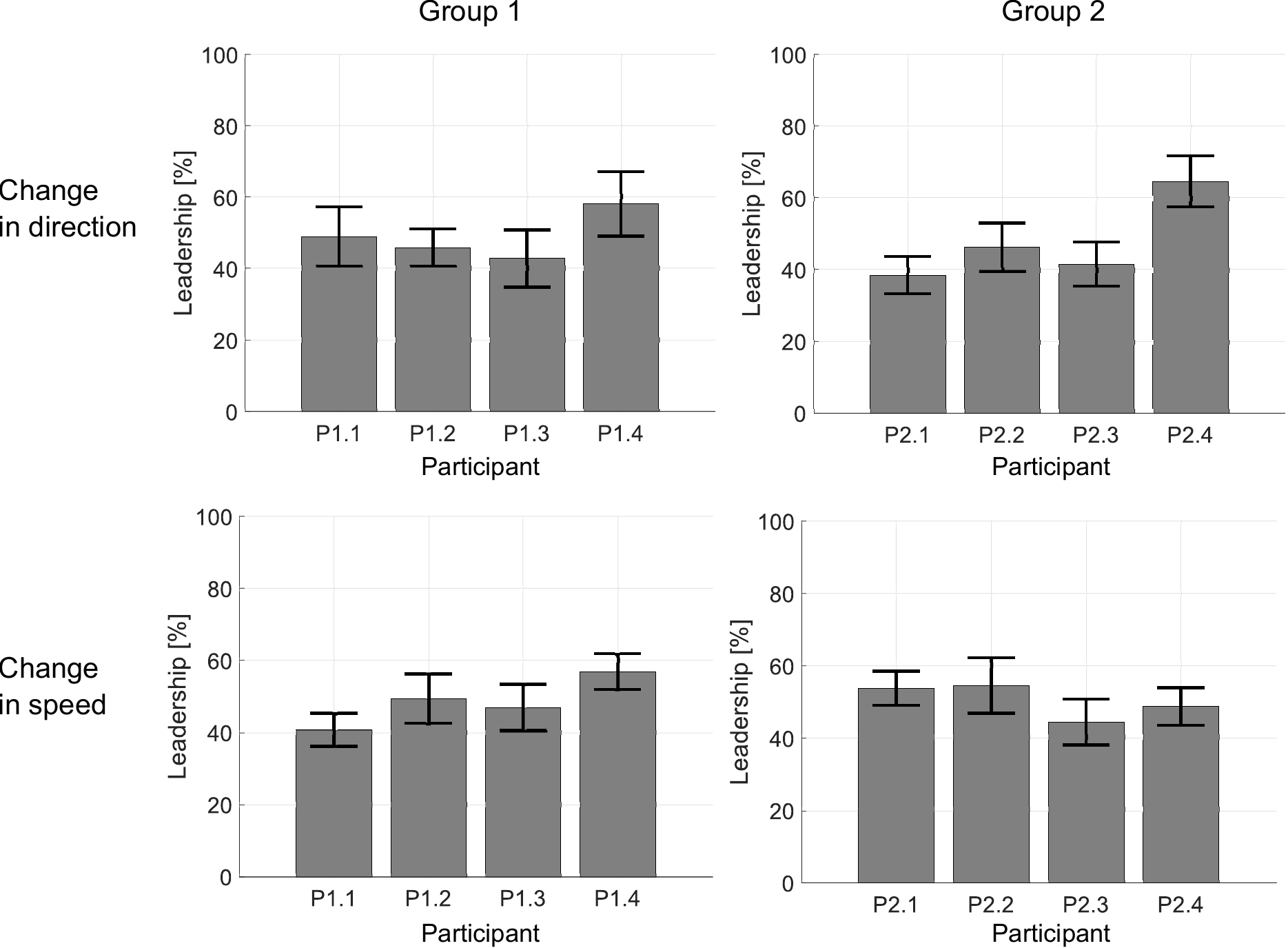}
	\caption{\textbf{Averaged percent leadership grouping the trials by individual pedestrian}. Four bar charts are reported, $2$ groups $\times$ $2$ condition (speed and heading direction). Each bar represents the mean value of the leadership grouped by a specific pedestrian over the number of trials independently from his/her position, whereas grey error bars are the standard error of the mean.}
	\label{fig:players_comparison}
\end{figure}

\subsection*{Effect of individual locomotor behaviour} The same data are replotted by participant in Figure \ref{fig:players_comparison}, where the mean value and standard error of the mean (SEM) of the individual leadership index are shown for each pedestrian. Notice that some participants have consistently higher indices than the others. A one-way ANOVA (4 Participants) on mean percent leadership was performed separately for each group revealing a main effect only in direction-change trials for Group $2$ ($F(3,24) = 4.663, p < 0.05, \eta^2 = 0.368$).

\begin{figure}[ht]
	\centering
	\includegraphics[width=0.8\linewidth]{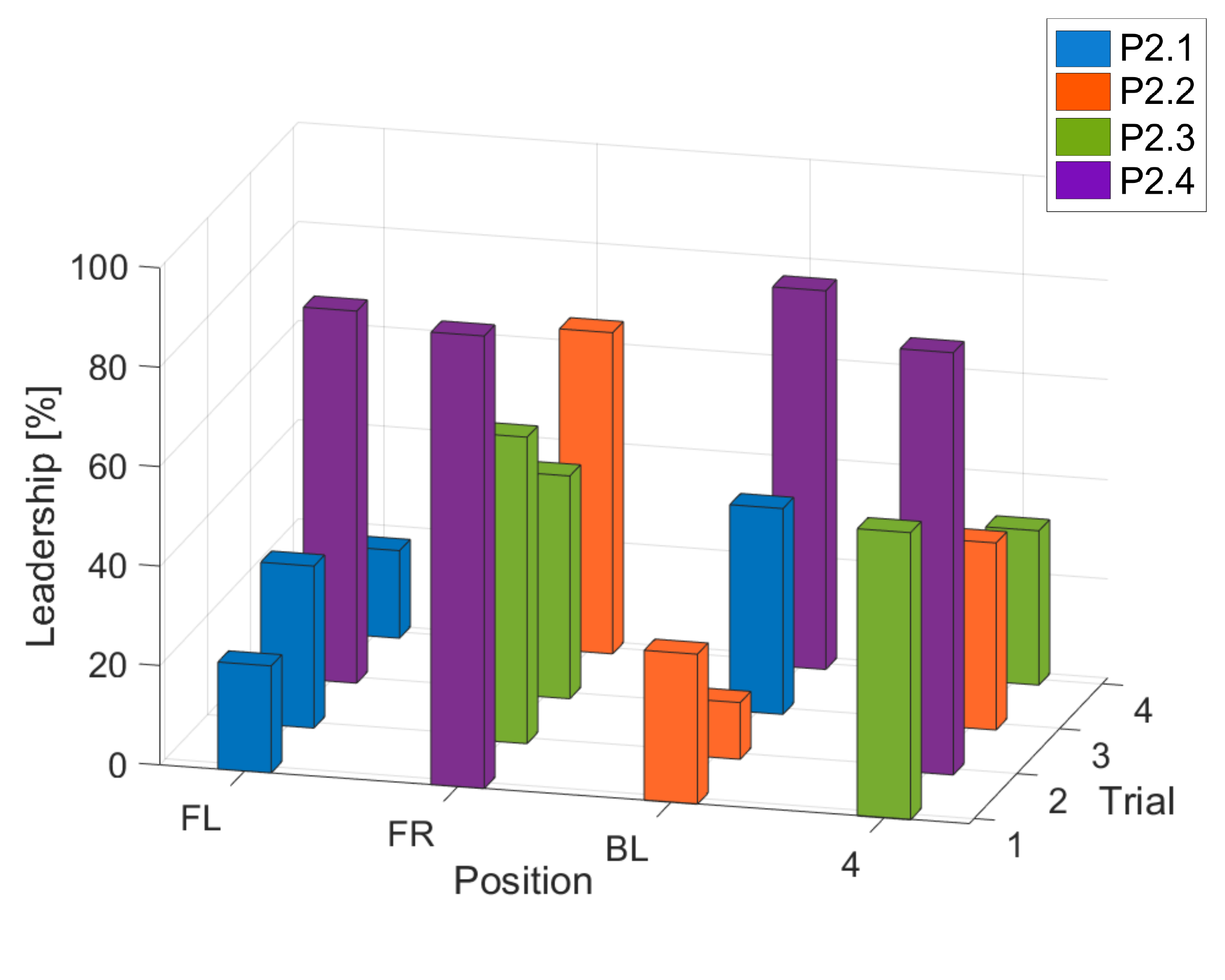}
	\caption{\textbf{Percent leadership of four sample trials having P$2.4$ in each position}. Percent leadership of four sample trials are reported having the pedestrian P$2.4$ (in purple) in each of position. The $x$-axis represents the trial, the $y$-axis the position (Front Left, Front Right, Back Left, Back Right), whereas the $z$-axis reports the value of leadership in percentage. Each pedestrian is identified by different colour.}
	\label{fig:absolute_leadership}
\end{figure}

In these trials, pedestrian P$2.4$ became the leader independently of his/her position in the group as shown in Figure \ref{fig:absolute_leadership} (details about leader-follower patterns inferred from the data are reported in \emph{Supplementary Info, Supplementary Figures 1,2,3 and 4}). Pairwise comparisons of participant means using a Bonferroni post-hoc test confirmed that the behaviour of pedestrian P$2.4$ has a significant influence on the behaviour of the whole group ($p < 0.05$). This influence is absent in the speed-change condition.

We attribute the leadership of pedestrian P$2.4$ to a characteristic behaviour, the tendency to initiate a change by acting first, which is evident in the time series (see \emph{Supplementary Info, Supplementary Figures 1,2,3 and 4}).  The question arises, however, as to how this individual could influence other group members from the back row when changing direction, but not when changing speed.  This discrepancy may be explained by the observation that on direction-change trials, P$2.4$ often moved into the peripheral field of view of other group members, whereas on speed-change trials, P$2.4$ remained directly behind the front row, out of view.

\section*{Discussion}
\label{sec:discussion}
In this work, we carried out a data-driven experimental investigation of leadership emergence in a group of walking pedestrians. Two groups of four volunteers each were asked to walk across a room together, and to change their heading or speed. We then analysed the acquired data on their movement and speed by defining a novel ``individual leadership index'' to identify the leading participant and reconstruct leader-follower interactions in the group. This metric is based on time-dependent correlation analysis and was adapted from the indexes \cite{Giuggioli2015} that were used previously to study the leader-follower behaviour of flying bats. To the best of our knowledge, it is applied here for the first time to a human group in order to predict leader-follower patterns.

Our analysis shows that both contextual factors, such as the position of an individual in the group, as well as personal factors, such as a characteristic locomotor behaviour, contribute significantly to the emergence of a leadership role.  In particular, our findings reveal that participants in the front row, to whom those in the back row are visually coupled, are usually elected as leaders via the mechanism of collective motion, in which pedestrians match the heading direction of neighbours in their field of view \cite{Rio2018}. However, we also find that a specific individual assumed a leadership role more than 60\% of the time, initiating a change in direction irrespective of their position in the group.  By deciding to turn first, this participant exploited the mechanism of collective motion to seize the leadership, a \textit{self-appointed} leader.

These findings demonstrate that leadership in walking groups is a complex phenomenon that does not only depend on geometric conditions, such as the leader location in the group, but also on other features related to the characteristics of individual participants; for example, the tendency to initiate movements or changes of direction that when perceived by neighbours propagate to the rest of the group. This observation might be relevant to understand how coordination in walking groups emerges in situations, such as emergency evacuations or some sport scenarios, where verbal communication among group members is limited.

We wish to emphasise that the present methodology reveals information about specific leader-follower relationships within a group and how they change over time, as illustrated in Figures \ref{fig:leadership_direction} and \ref{fig:leadership_speed}.  Moreover, the method can be generalised to larger pedestrian groups and other scenarios.  An understanding of leadership emergence and the factors on which it depends can be extremely useful in emergency situations when it is critical to influence group dynamics, such as guiding a crowd to a safe exit.

\section*{Methods}
\label{sec:methods}

\subsection*{Experiment}
\paragraph{Participants.}
Two groups of $4$ participants were tested in separate sessions as part of a larger study. The research protocol was approved by Brown University's Institutional Review Board, in accordance with the Declaration of Helsinki. Informed consent was obtained from all participants, who were compensated for their time. All participants had normal or corrected vision, and none reported a motor impairment.

\begin{figure}[ht]
	\centering
	\includegraphics[width=0.7\linewidth]{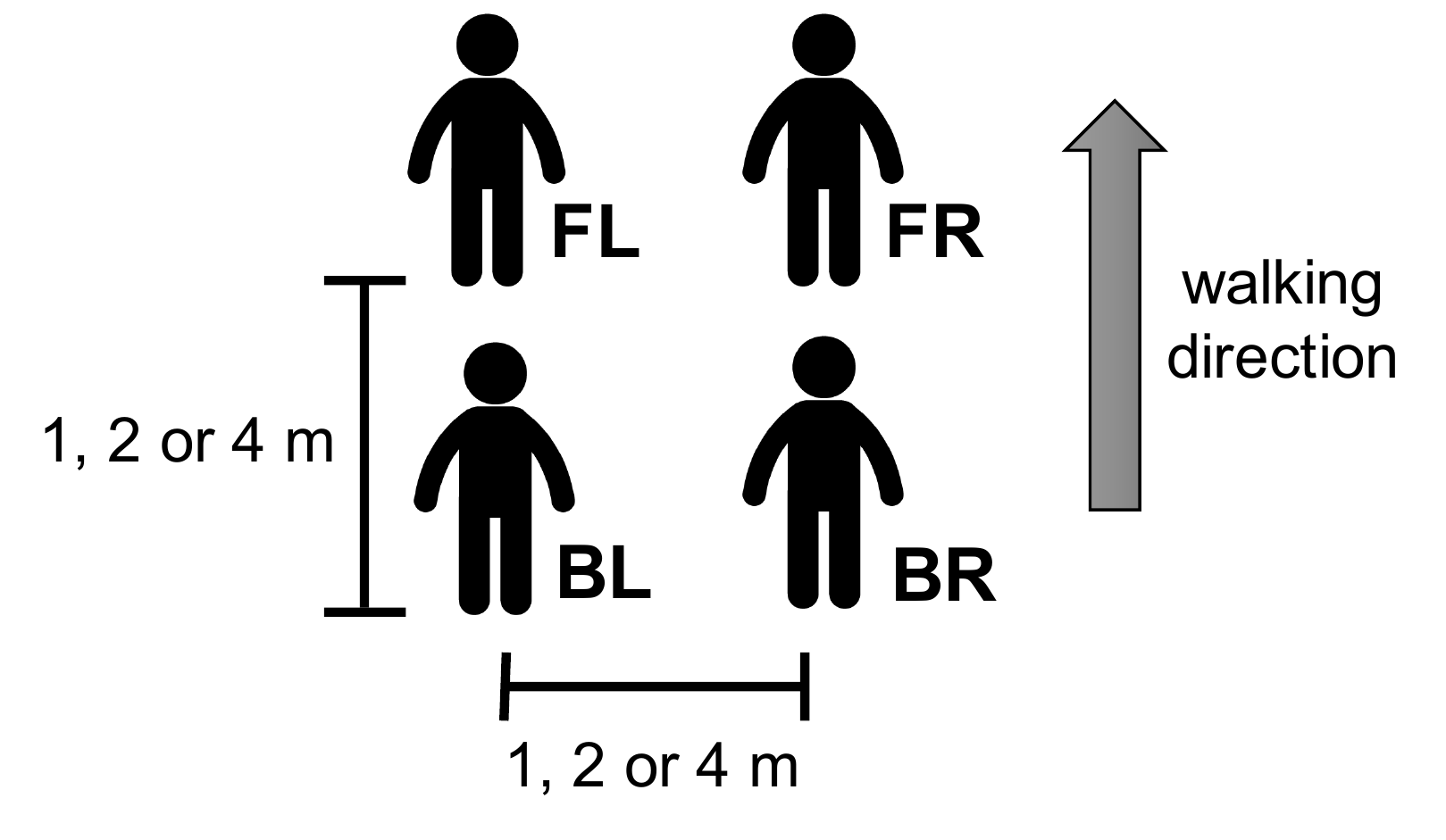}
	\caption{\textbf{Group configuration.} Participants are placed at the vertices of a square (FL, FR, BL, BR) with an initial inter-personal distance (IPD) between them of $1$, $2$ or $4$ metres.}
	\label{fig:group_config}
\end{figure}

\paragraph{Task and procedure.}
Before each trial, participants were positioned at the vertices of a square configuration marked on the floor (``front left'' (FL), ``front right'' (FR), ``back left'' (BL), ``back right'' (BR)), at one of three initial inter-personal distances (IPD of $1$, $2$, or $4$m) (see Figure \ref{fig:group_config}). They were instructed to change direction or speed twice at self-selected times as they walked together across the room (about $20$m), and to ``try to stay together with your neighbours''. There were nine possible instructions: change direction in left/right sequence (LL, RR, LR, or RL), change speed in low/fast sequence (SS, FF, SF, or FS), or no-change control (CC). The experimental design was thus $2$ Conditions (heading direction, speed) $\times$ $4$ Sequences $\times$ $3$ IPD ($1$, $2$, $4$ m), with one trial in each of the $24$ conditions plus three control trials. Given that no effects of IPD were found, data were collapsed across sequence and IPD for analysis, yielding $12$ trials per analysed condition.

At the beginning of each trial, a sequence instruction was given, then the group started walking upon a verbal ``ready, begin'' command.  Before the next trial, participants shuffled their positions in the starting configuration.  Each group received one practice trial, followed by a block of $12$ heading change trials, and a block of $12$ speed change plus $3$ control trials, for a total of $27$ test trials.  Trial order was randomised within blocks, and block order was counterbalanced across the two groups. A test session lasted about $40$ min. Further details about each single trial are reported in \textit{Supplementary Info, Supplementary Tables 1 and 2}.

\subsection*{Data acquisition and analysis}
The experiments were carried out at Brown University in a large hall with a $14$m $\times$ $20$m tracking area marked on the floor. Head position was recorded at $60$ Hz with a $16$-camera infrared-reflective motion capture system (Qualisys Oqus). Each participant wore a bicycle helmet having $5$ markers placed on $30-40$cm stalks in a unique configuration, so that each helmet could be identified.

The time series of head positions of all four pedestrians were recorded in three dimensions, but only two dimensions ($xy$-plane) were analysed further. Time series were filtered using a forward and backward $4$th-order low-pass Butterworth filter to reduce tracker error and gait oscillations, before differentiating to obtain time series of speed and heading. A $1.0$ Hz cutoff was used to reduce anterior-posterior oscillations on each step before computing speed, and a $0.6$ Hz cutoff for reducing medial-lateral oscillations on each stride before computing heading. Files were truncated to eliminate endpoint error. Leadership was then evaluated by assessing the difference in heading direction at each time between each pair of pedestrians by adapting the time-dependent delayed directional correlation $C^d$ (see \emph{Leadership metrics}) \cite{Giuggioli2015}. Summing the percentage of time in which a specific pedestrian is identified as leader, we were able to reconstruct a network representing the influence that a specific pedestrian has on any other. Similarly, leadership in speed was evaluated by adapting the previous algorithm using time-dependent delayed speed correlation $C^s$ instead.

\subsection*{Leadership metrics}
\label{sec:leadership_metrics}
\paragraph{Time-dependent delayed directional correlation (TDDC).}
Using time-dependent delayed directional correlation (TDDC), we characterised leadership in the walking group in terms of the heading direction of each individual, spotting whether some of the participants adjusted their own direction to that of the others and exhibited a follower role. Accordingly, a leader is defined as the participant whose heading direction influences the others the most, in the sense that the other participants match their headings with that of the leader after a time delay corresponding to the natural human visual-locomotor delay.

Denoting by $v_k \in R^2$ the velocity vector of   pedestrian $k$ in the plane, we compute:
\begin{equation}
h_{ij}(t,\tau) = \dfrac{v_i(t)}{||v_i(t)||} \dfrac{v_j(t+\tau)}{||v_j(t+\tau)||} \quad \in [-1, 1],
\end{equation}
which represents the scalar product between the heading direction of pedestrian $i$ at time $t$ and that of pedestrian $j$ at delayed time $t+\tau$.

Formally, TDDC is defined as follows:
\begin{equation}
C^d_{ij}\left(t,\tau\right) = \frac{1}{2\omega + 1}\sum_{k=-\omega}^{\omega}h_{ij} \left(t+k\Delta t, \tau\right),
\end{equation}
where $\Delta t$ is the sampling time and $\omega > 0$ is a suitably selected constant. In order to compensate for high-frequency noise and measurement errors, $h_{ij}(t,\tau)$ is averaged over a symmetric time-window $2 \omega \Delta t$. As the measured human visual-locomotor delay to the leader's change in heading is approximately equal to $1000$ ms and the sampling time for the collected data is equal to $\Delta t = \dfrac{1}{60 Hz}$, we have that $2 \omega \Delta t > 1s$ and so $\omega$ was set to $40$.

To identify the leader, $C^d_{ij}(t, \tau)$ was evaluated for each pair of pedestrians $(i,j)$ at every time instant $t$ along the trial and for different values of delay $\tau$. If the value of $\tau_{ij}(t)$ maximising the heading direction correlation is positive then we say that pedestrian $i$ leads pedestrian $j$ at time instant $t$, or otherwise if that value is negative.

\begin{figure}[!ht]
	\centering 
	\includegraphics[width=0.93\textwidth]{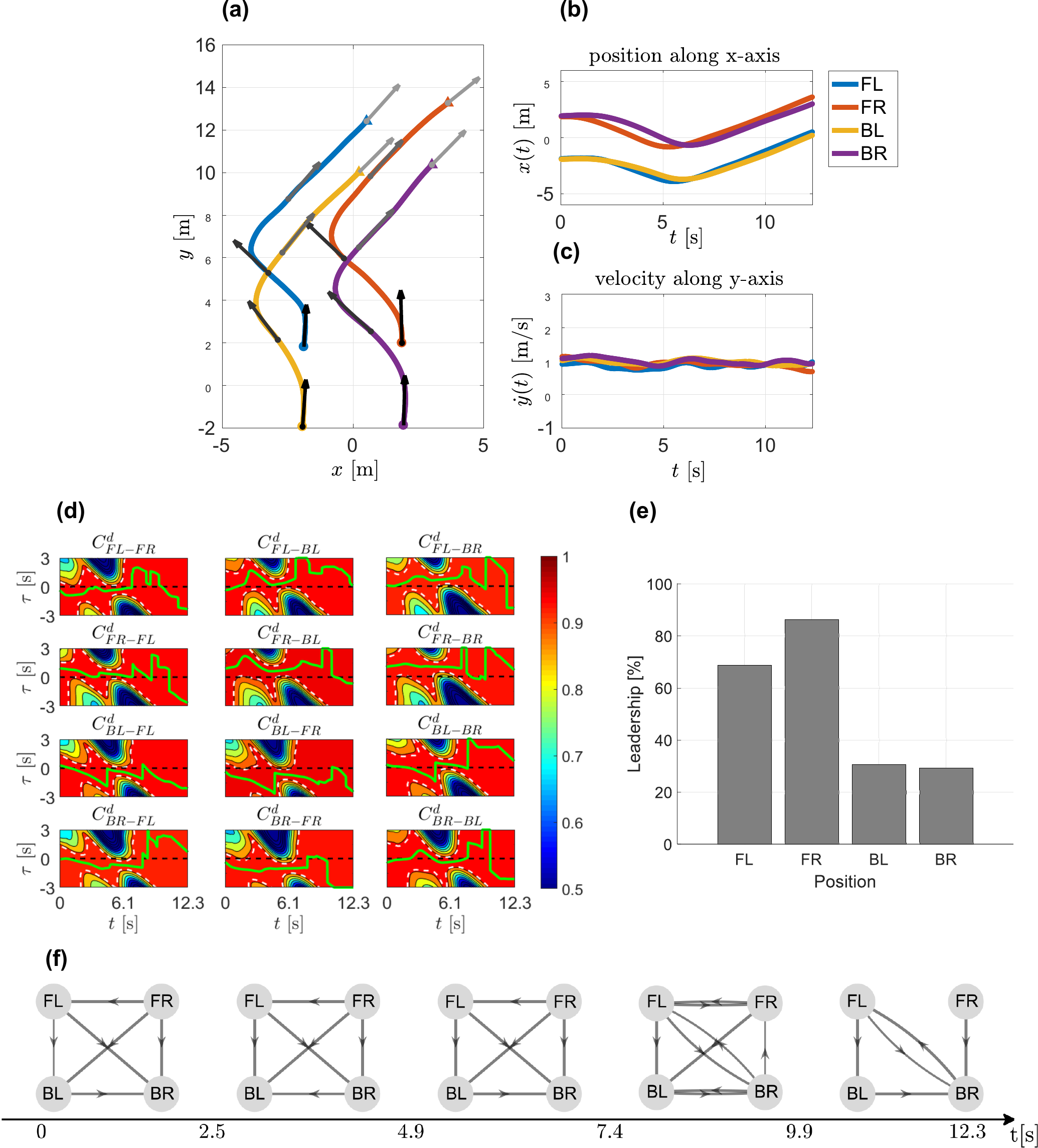}
	\caption{\textbf{Left-right representative trial}. \textbf{(a)} Pedestrians' trajectories in the $xy$-plane with different colours. Velocity vectors are represented by an arrow each $3$ seconds, the colour goes from black to grey as time increases.\textbf{(b-c)} Position along $x$-axis and velocity along $y$-axis. Pedestrians are identified by his/her position (FL, FR, BL, BR). \textbf{(d)} TDDC evaluated for each pair of pedestrians. The $x$-axis represents time instant $t$, whereas the $y$-axis represents different values of the time-delay $\tau$. Different colours are different values of TDDC, with blue representing $C^d_{ij}(t,\tau)\le 0.5$ and red representing $C^d_{ij}(t,\tau)=1$. Dashed white lines represent level curves where $C^d_{ij}(t,\tau)=0.95$, whereas the dashed black line represents $\tau(t)=0$. The solid green line on top of the heat map represents the value of $\tau(t)$ maximising the heading direction correlation between two pedestrians. \textbf{(e)} Percent leadership. Each grey bar refers to a different member in the group. \textbf{(f)} Network reconstruction. Each node represents a pedestrian, whereas the arrow directed from node $i$ to node $j$ indicates the fact that the heading direction of pedestrian $i$ influences that of pedestrian $j$. The arrow's width is the strength of such interaction.}
	\label{fig:leadership_direction}
\end{figure}

A left-right trial carried out by Group $1$ with an initial IPD of $4$m is taken as a representative example to illustrate the analysis based  on time-dependent directional correlation metric. Figure \ref{fig:leadership_direction}(d) shows the heat maps characterising values of $C^d_{i,j}(t, \tau)$ for the whole duration of the trial with different time delay $\tau$ between each pair ($i$, $j$) of pedestrians.

\paragraph{Time-dependent delayed speed correlation (TDSC).}
To assess leadership roles in speed-change trials, we define:

\begin{equation}
\Delta s_{ij}(t, \tau) = \left| \; ||v_i(t)|| - ||v_j(t+\tau)|| \; \right|
\end{equation}
which represent the difference between the speed of pedestrian $i$ at time $t$ and that of pedestrian $j$ at delayed time $t+\tau$. The time-dependent speed correlation (TDSC) is defined as:
\begin{equation}
\centering
C^s_{ij}\left(t,\tau\right) = \frac{1}{2\omega + 1}\sum_{k=-\omega}^{\omega}\Delta s_{ij} \left(t+k\Delta t, \tau\right),
\end{equation}
where the parameter $\omega$ is the length of the symmetric window used to average $s_{ij}(t,\tau)$ in order to compensate for noise and measurements errors as done before. As the measured human visual-locomotor delay to the leader's change in speed is approximately equal to $500$ ms \cite{Rio2014b}, we have set $\omega$ to $20$. In this case pedestrian $i$ is said to lead pedestrian $j$ at time $t$ if the value of $\tau_{ij}(t)$ minimising the speed correlation is positive, or otherwise if it is negative.

\begin{figure}[!ht]
	\centering
	\includegraphics[width=0.93\textwidth]{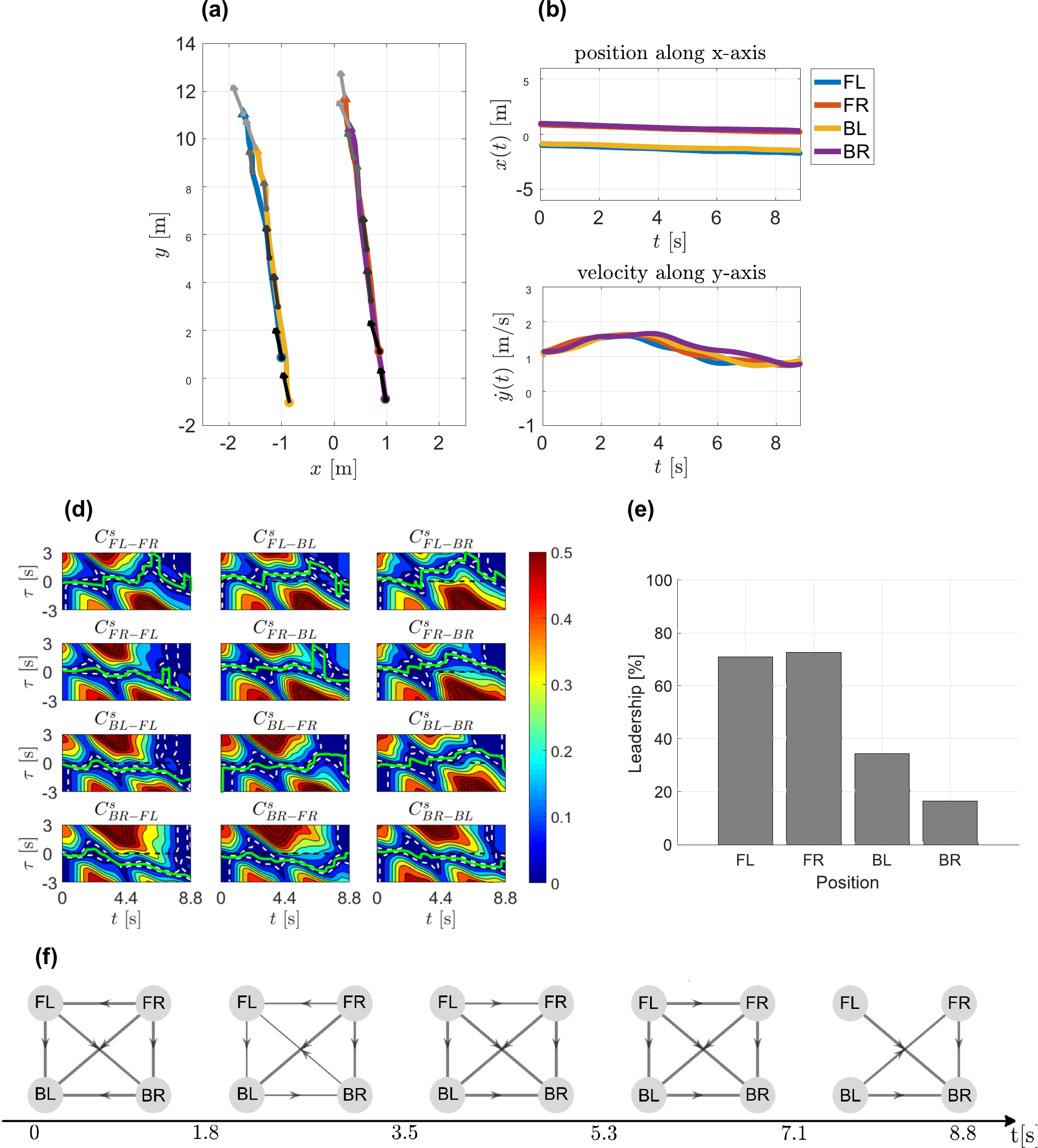}
	\caption{\textbf{Fast-slow representative trial}. \textbf{(a)} Pedestrians' trajectories in the $xy$-plane with different colours. Velocity vectors are represented by an arrow each $3$ seconds, the colour goes from black to grey as time increases.\textbf{(b-c)} Position along $x$-axis and velocity along $y$-axis. Pedestrians are identified by his/her position (FL, FR, BL, BR). \textbf{(d)} TDSC evaluated for each pair of pedestrians. The $x$-axis represents time instant $t$, whereas the $y$-axis represents different values of the time-delay $\tau$. Different colours are different values of TDSC, with blue representing $C^s_{ij}(t,\tau)=0$ and red representing $C^s_{ij}(t,\tau)\ge 0.5$. Dashed white lines represent level curves where $C^s_{ij}(t,\tau)=0.05$, whereas the dashed black line represents $\tau(t)=0$. The solid green line on top of the heat map represents the value of $\tau(t)$ minimising the speed correlation between two pedestrians. \textbf{(e)} Percent leadership. Each grey bar refers to a different member in the group. \textbf{(f)} Network reconstruction. Each node represents a pedestrian, whereas the arrow directed from node $i$ to node $j$ indicates the fact that the speed of pedestrian $i$ influences that of pedestrian $j$. The arrow's width is the strength of such interaction.}
	\label{fig:leadership_speed}
\end{figure}

A fast-slow trial carried out by group $1$ with an initial IPD of $2$m is taken as a representative example to illustrate the analysis of the data based on time-dependent speed correlation metric. Figure \ref{fig:leadership_speed}(d) shows the heat maps characterising values of $C^s_{i,j}(t, \tau)$ for the whole duration of the trial with different time delay $\tau$ between each pair ($i$, $j$) of pedestrians.

\paragraph{Individual leadership index.}
Knowledge of $\tau_{ij}(t)$, maximising/minimising the heading direction/speed correlation, for each trial and pair of pedestrians allows to quantitatively estimate how much a participant acts as a leader in the group. Specifically we define the {\em individual leadership index} for each participant as the overall average percentage of time for which the computed $\tau_{ij}(t)$ is positive (Figure \ref{fig:leadership_direction}(e) and Figure \ref{fig:leadership_speed}(e)).

\paragraph{Network reconstruction.}
Knowledge of $\tau_{ij}(t)$ allows also to infer leader-follower interactions between each pair of group members. Each pedestrian is represented as a node in the network and the influence that each pedestrian has on the others is represented as an edge (an edge directed from node $i$ to node $j$ means that pedestrian $j$ is influenced by pedestrian $i$). The strength of this influence is captured by the edge weight $w_{i,j} \in [0,1]$ representing the normalised percentage of time in which pedestrian $i$ leads pedestrian $j$. The network were reconstructed repeatedly within $5$ consecutive time windows in order to evaluate also the evolution of the interactions over time (Figure \ref{fig:leadership_direction}(f) and Figure \ref{fig:leadership_speed}(f)).

The topologies inferred were then refined by means of DPI techniques (data processing inequality) and by a further thresholding to remove false positive links between nodes, as suggested in \cite{Alderisio2017}. Specifically: i) links whose weights $w_{i,j}$ are null were discarded from the reconstructed network, and ii) non-zero weights $w_{i,j}$ were checked among all triplets of connected nodes so that, on the basis of their intensity, one of the triplet can be possibly regarded as a false connection and set to zero (so that the corresponding link was removed). For example, let us consider three nodes $i$, $j$ and $k$. If the weight $w_{i,k}$ between the nodes $i$ and $k$ was lower than $w_{i,j}$ between the nodes $i$ and $j$, and was also lower than $w_{j,k}$ between the nodes $j$ and $k$, then the weight $w_{i,k}$ was set to zero and the link between the two nodes $i$ and $k$ was removed.

\newpage

\newpage

\bibliography{library}

\section*{Acknowledgements}
This research was supported by R01EY010923 from the National Institutes of Health (USA), and BCS-1849446 from the National Science Foundation (USA), and EPSRC PhD Scholarship to M.L. The authors wish to thank Kevin Rio, Zach Page, Stephane Bonneaud, Adam Kiefer, Michael Fitzgerald, and the Sayles Swarm team for help with data collection, and Greg Dachner and the VENLab crew for assistance with data processing.

\section*{Author contributions statement}
Conceived and designed the experiments: W.H.W. Performed the experiments and collected the data: W.H.W. Metrics definition and performed the data analysis: M.L., M.d.B. Wrote the paper: M.L., M.d.B., W.H.W. All authors gave final approval for publication.

\section*{Additional information}
\textbf{Supplementary information} accompanies this paper; \textbf{Competing Interests}: The authors declare that they have no competing interests.

\end{document}